\documentclass[aps,twocolumn,pra,showpacs,superscriptaddress,10pt]{revtex4-1}

\usepackage{amssymb}
\usepackage{amsmath}
\usepackage{graphicx,color}
\usepackage{epstopdf}
\usepackage{mathrsfs}
\usepackage{epsfig}
\usepackage[breaklinks=true]{hyperref}

\newcommand{\beq}{\begin{equation}}
\newcommand{\eeq}{\end{equation}}
\newcommand{\bqa}{\begin{eqnarray}}
\newcommand{\eqa}{\end{eqnarray}}

\definecolor{green}{rgb}{0.00,0.50,0.00}

\begin{document}

\title{On The Scattering Process in Quantum Optics}
\author{John E.~Gough} \email{jug@aber.ac.uk}
\affiliation{Aberystwyth University, Aberystwyth, SY23 3BZ, Wales, United Kingdom}
\date{\today}

\begin{abstract}
The derivation of a quantum Markovian model for an opto-mechanical system consisting of a 
quantum mechanical mirror interacting with quantum optical input fields via radiation pressure
is difficult problem which ultimately involves the scattering process of
quantum stochastic calculus. We show that while the scattering process may
be approximated in a singular limit by regular processes using different
schemes, however the limit model is highly sensitive to how the approximation scheme is
interpreted mathematically. We find two main types of stochastic limits of regular models, and 
illustrate the origin of this difference at the level of one particle scattering.
As an alternative modelling scheme, we consider models of mirrors as non-trivial dielectric medium
with a boundary that is itself quantized. Rather than treating the plane waves for the electromagnetic
field, we take the actual physical modes and quantize these. The input-output formalism
is then obtained in the far zone where the plane wave approximation is valid. Several examples
are considered, and the quantum stochastic model is derived. We also consider the quantum trajectories
problem for continual measurement of the reflect output fields, and derive the stochastic
master equations for homodyning and photon counting detection to estimate the mirror observables.
\end{abstract}

\pacs{03.65.Nk, 42.50.Dv, 03.65.Db, 02.30.Mv}

\maketitle

\section{Introduction}

The theory of quantum stochastic calculus of Hudson and Parthasarathy \cite{HP} has a
long history of applications to quantum open systems. In addition to
introducing quantum stochastic integrals with respect to creation and
annihilation processes, they included integration with respect to scattering
processes. While not originally motivated by quantum input-output models,
this additional feature allowed unitary rotations of input fields to be
considered in conjunction to displacements. This has lead to a unified
treatment of state-based input-output models that has been exploited in
model interconnections of such models into quantum feedback networks, and in
particular to rigorously model simplification and reduction procedures such
as adiabatic elimination of open systems \cite{AE}-\cite{GN}; other areas of
application include the qubit limit of cQED \cite{Mabuchi}.

The scattering operator $S$ appearing in the unitary quantum stochastic differential equations (QSDE)
however has attracted some comparison to  
$S$-matrix of usual scattering theory, however, its origin and role is rather different. The scattering processes are also 
less familiar to the Phsyics community than the usual creation annihilation processes, as they where not part of the original
input/output formalism of quantum optics \cite{ColGar84},\cite{GarCol85}.

Typically quantum stochastic processes obeying a non-trivial It\={o} table
arise as singular limits of approximating regular processes. This is a
delicate problem for classical stochastic processes, however there are
additional issues related to approximating the scattering processes as
different schemes for otherwise similar dynamics lead to different limit
evolutions. 

For an $m$ input model we have quantum white noise input process $b_{j}( t) $
for $j=1,\cdots ,m$ satisfying singular commutation relations $[ b_{j}( t)
,b_{k}^{\dag }( s) ] =\delta _{jk}\delta ( t-s) $ and we define the $( m+1)
^{2}$ fundamental processes \cite{HP} 
\begin{eqnarray}
\Lambda _{\alpha \beta }( t) =\int_{0}^{t}b_{\alpha }^{\dag }( s) b_{\beta
}( s) ds
\end{eqnarray}
where we also include the index 0 by setting $b_{0}( t) \equiv 1$. In this
way $\Lambda _{00}( t) =t$, while $B_{j}( t) :=\Lambda _{0k}( t)
=\int_{0}^{t}b_{k}( s) ds$ and $B_{j}^{\dag }( t) :=\Lambda _{j0}( t)
=\int_{0}^{t}b_{j}^{\dag }( s) ds$ are the processes of annihilation and
creation. $\Lambda _{jk}( t) $ describes the process where a quanta in
channel $k$ is annihilated and another immediately created in channel $j$ at
some time in the interval $[ 0,t] $. The $\Lambda _{\alpha \beta }( t) $ are
well-defined operators acting on the Fock space $\mathfrak{F}$ over $\mathbb{C}%
^{m}$-valued square-integrable functions of positive time $t\geq 0$. We note
the quantum It\={o} table \cite{HP} 
\begin{eqnarray}
d\Lambda _{\alpha \beta }( t) d\Lambda _{\mu \nu }( t) \equiv \hat{\delta}%
_{\beta \mu }d\Lambda _{\alpha \nu }( t)
\end{eqnarray}
where $\hat{\delta}_{\beta \nu }=1$ if $\beta =\nu \neq 0$, and vanishes
otherwise.

Let us fix a system with Hilbert space $\mathfrak{h}$. The quantum stochastic
differential equation (QSDE) on $\mathfrak{h}\otimes \mathfrak{F}$ (implied sum over
repeated Greek indices from 0 to $m$) 
\begin{eqnarray}
dU(t)=G_{\alpha \beta }\otimes d\Lambda _{\alpha \beta }(t)\,U(t),
\label{eq:QSDE}
\end{eqnarray}
with initial condition $U(0)=I$, possesses a unique solution for bounded
operators $G_{\alpha \beta }$ on $\mathfrak{h}$. The necessary and sufficient
conditions for the process $U(t)$ to be unitary are that (implied sum over
repeated Latin indices from 1 to $m$) \cite{HP} 
\begin{eqnarray}
G_{jk} &=&S_{jk}-\delta _{jk},G_{j0}=L_{j},  \notag \\
G_{0k} &=&-L_{l}S_{lk},G_{00}=-\frac{1}{2}L_{l}^{\dag }L_{l}-iH
\label{eq:SLH}
\end{eqnarray}
with $S=[S_{jk}]$ unitary, $L=[L_{j}]$ arbitrary, and $H$ self-adjoint. The
triple $(S,L,H)$ then determines the model.

\subsection{Approximations by Regularized Hamiltonians}
In practice, the singular processes are idealisations. However, working in the same Fock space, it is possible
to approximate the fundamental processes by regular ones obtained by
smearing with some mollifying function. In the following, we will denote by $\delta _n$ a regular function with compact support parametrized by $n>0$
and converging to a delta function as $n\rightarrow \infty $. For
definiteness we may fix an integrable function $g$ with support $[-c,c]$ (i.e., $g(x)=0$ for $|x|>c$) for
some finite range $c>0$, such that $g(-x)=g(x)$ and $\int_{-c}^{c}g(x)dx=1$. We may then take $\delta _{n}(x)=n\, g(nx)$, and these
vanish outside the interval $[-c/n,c/n]$.

We shall now describe two possible schemes to formally approximate $b_{\alpha}^\dag ( t) b_{\beta }( t) $. 

Let $E_{\alpha \beta }$ be a collection of
operators on the Hilbert space $\mathfrak{h}$ of a fixed system such that $E_{\alpha \beta }^\dag =E_{\beta \alpha }$. 
For convenience we assume that
they are bounded. We write $E_{\ell \ell }$ for the matrix $[ E_{jk}] $, $E_{\ell 0}$ for the column vector $[ E_{j0}] $, and $E_{0\ell }$ for the row
vector $[ E_{0k}] $.

The matrix $E_{\ell \ell}$ will be called the \textit{exchange matrix}. In principle, $E_{jk}$ gives the strength of the interaction causing an input quantum
of type $k$ to be annihilated and replace with a quantum of type $k$.

\subsubsection{Scheme \#1}

We set $\tilde{\lambda}_{\alpha \beta }^{( n) }( t) =\int \delta _{n}(
t-s) b_\alpha ^\dag ( s) b_\beta ( s) ds$, or more exactly
\begin{eqnarray}
\tilde{\lambda}_{\alpha \beta }^{( n) }( t) &=&\int \delta _{n}( t-s)
d\Lambda _{\alpha \beta }( s) , 
\end{eqnarray}
\begin{eqnarray}
\tilde{b}_{k}^{( n) }( t) &=&\int \delta _{n}( t-s) dB_{k}( s) .
\end{eqnarray}

A unitary $\tilde{U}^{( n) }( t) $ is defined as the solution to the
Schr\"{o}dinger equation with time-dependent Hamiltonian 
\begin{eqnarray}
\tilde{H}^{( n) }(t) &=& \int \delta _{n}( t-s) E_{\alpha \beta} \otimes
d\Lambda _{\alpha \beta }( s) , \nonumber \\
&=& E_{\alpha \beta }\otimes \tilde{\lambda}_{\alpha \beta }^{( n) }( t) \nonumber \\
&=& E_{jk }\otimes \tilde{\lambda}_{jk }^{( n) }( t) \nonumber \\
&+& E_{j0 }\otimes \tilde{b}_{j }^{( n) \dag}( t) +E_{0k }\otimes \tilde{b}_{k }^{( n) }( t) +E_{00}.
\end{eqnarray}

\begin{widetext}
The limit process $\tilde{U}( t) $
then exists, is unitary and described by the triple 
\begin{eqnarray}
\tilde{S}=e^{-iE_{\ell \ell }},\quad \tilde{L}=\frac{e^{-iE_{\ell \ell }}-1}{%
E_{\ell \ell }}E_{\ell 0},\quad \tilde{H}=E_{00}-E_{01}\frac{E_{\ell \ell
}-\sin ( E_{\ell \ell }) }{( E_{\ell \ell }) ^{2}}%
E_{\ell 0}.
\end{eqnarray}
The limit is best understood as a trotterized time-ordered exponential introduced by Holevo \cite{Holevo}
\begin{eqnarray}
\tilde{U}( t) =\vec{T}_{H}e^{-i\int_{0}^{t}dE}:=\lim_{\max |
t_{k+1}-t_{k}| \to 0}e^{-iE( t_{N}, t_{N-1}) }\cdots
e^{-iE( t_{2},t_{1}) }e^{-iE( t_{1},t_{0}) }
\label{eq:Holevo}
\end{eqnarray}
where $E( t_{2},t_{1}) =\int_{t_{1}}^{t_{2}}E_{\alpha \beta
}\otimes d\Lambda _{\alpha \beta }( s) \equiv E_{\alpha \beta
}\otimes \{ \Lambda _{\alpha \beta }( t_{2}) -\Lambda
_{\alpha \beta }( t_{1}) \} $ and $t=t_{N}>\cdots
>t_{1}>t_{0}=0$. 
\end{widetext}

The relationship between the coefficients is $\tilde{G}%
_{\alpha \beta }\otimes d\Lambda _{\alpha \beta }\equiv e^{-i E_{\alpha \beta
}\otimes d\Lambda _{\alpha \beta }}-1$. 

\subsubsection{Scheme \#2}

We alternatively set 
\begin{eqnarray}
\lambda _{jk}^{( n) }( t) =\tilde{b}_{j}^{( n) \dag }( t) \tilde{b}_{k}^{(
n) }( t)
\end{eqnarray}
and $\lambda _{j0}^{( n) }( t) =\tilde{b}_{j}^{( n) \dag }( t) ,\lambda
_{0k}^{( n) }( t) =\tilde{b}_{k}^{( n) }( t) $, $\lambda _{00}^{( n) }( t)
=1 $.

A unitary $U^{( n) }( t) $ is defined as the solution to the Schr\"{o}dinger
equation with time-dependent Hamiltonian 
\begin{eqnarray}
H^{( n) }( t) &=& E_{\alpha \beta
}\otimes \lambda _{\alpha \beta }^{( n) }( t) \nonumber \\
&=& E_{jk }\otimes \tilde{b}_{j }^{( n) \dag}( t) \tilde{b}_{j }^{( n) }( t) \nonumber \\
&+& E_{j0 }\otimes \tilde{b}_{j }^{( n) \dag}( t) +E_{0k }\otimes \tilde{b}_{k }^{( n) }( t) +E_{00}
.
\end{eqnarray}
The limit process $U( t) $
exists and is then described by the triple \cite{WZ}

\begin{widetext}
\begin{eqnarray}
S=\frac{I-\frac{i}{2}E_{\ell \ell }}{I+\frac{i}{2}E_{\ell \ell }},\quad L
=\frac{-i}{I+\frac{i}{2}E_{\ell \ell }}E_{\ell 0},\quad H=E_{00}+\frac{1}{2}
E_{0 \ell}\text{Im}\{ \frac{I}{I+\frac{i}{2}E_{\ell \ell }}\} E_{\ell
0}.
\end{eqnarray}
The limit process is best understood as the Stratonovich (or symmetric)
integral \cite{Stratonovich}
\begin{eqnarray}
U( t) =1-i\int_{0}^{t}dE( s) \circ U( s)
\equiv 1-i\lim_{\max | t_{k+1}-t_{k}| \to
0}\sum_{k}E( t_{k+1},t_{k}) U( \frac{t_{k+1}+t_{k}}{2}%
) 
\end{eqnarray}
using a midpoint rule for sample the integrand.

\end{widetext}

The Stratonovich unitary may be denoted as
\begin{eqnarray}
U(t) =\vec{T}_{D}e^{-i\int_{0}^{t}dE}
\end{eqnarray}
and formally the ordering $\vec{T}_D$ is the Dyson chronological ordering of the 
white noise operators $b_j (t)$ and $b_j^\dag (t)$, \cite{WZ}

\section{Single Particle Scattering Models}
\label{sec:single}
\begin{figure}[htbp]
	\centering
		\includegraphics[width=0.40\textwidth]{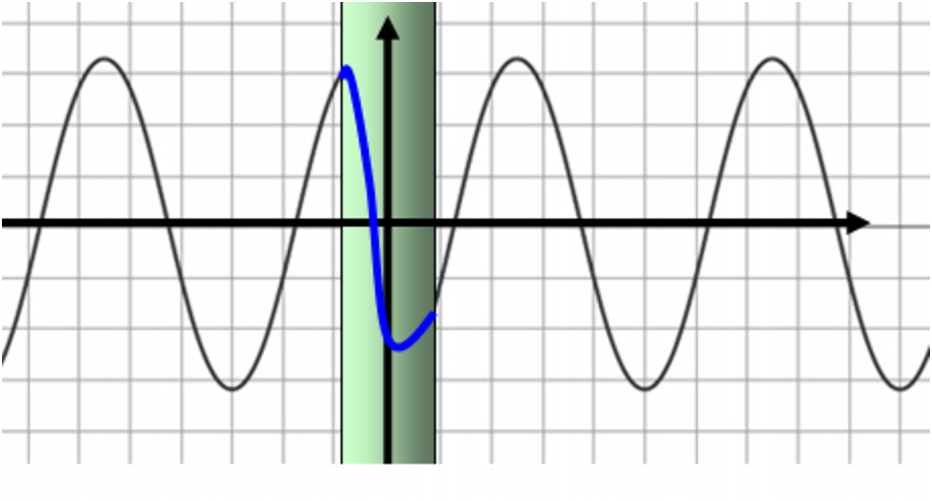}
	\caption{(Color online) A solution to (\ref{eq:H_reg}) with a region about the origin where $V \neq0$,
	and plane wave behaviour outside.}
	\label{fig:jump_reg}
\end{figure}

The limit procedures above are technically involved, however, we can obtain
some insight into what is going on at a simpler level. We consider the
situation of a quantum particle moving along the $x$-axis with Hamiltonian 
\begin{eqnarray}
H=-p+V
\label{eq:H_reg}
\end{eqnarray}
The free part $-{p}=i\partial $ described propagation at unit speed down
the axis, while the potential ${V}$ is localized in some region about
the origin, see Fig. \ref{fig:jump_reg}.

\begin{figure}[htbp]
	\centering
		\includegraphics[width=0.40\textwidth]{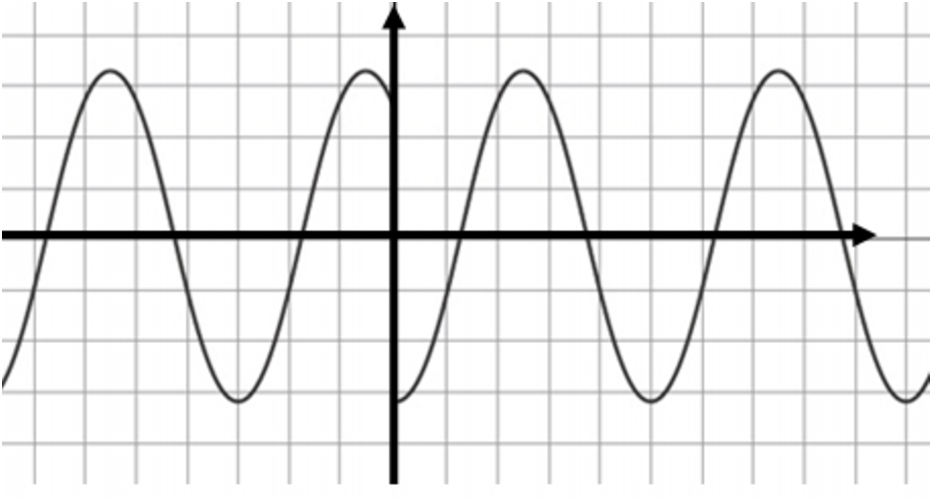}
	\caption{(Color online) The limit of a singular potential. A plane wave solution with a phase jump $s$ at the boundary.}
	\label{fig:jump_delta}
\end{figure}

If the potential is to be modeled as exactly localized at the origin - say a
delta potential ${V}=\varepsilon \delta ( {x}) $. then we are forced
to consider the operator $i\partial $ for $x\neq 0$ but this is not a
self-adjoint operator. In this case the corresponding ${H}$ must be a
self-adjoint extension of $i\partial $ on the punctured line and it is well
known that its domain is the set of all functions $\psi $ with derivative $%
\psi ^{\prime }( x) $ well-defined for $x\neq 0$ and $( \int_{-\infty
}^{0}+\int_{0}^{-\infty }) ( | \psi ( x) | ^{2}+| \psi ^{\prime }( x) |
^{2}) dx<\infty $ satisfying a boundary condition (see section X.1 of Reed
and Simon \cite{RSv2}, volume 2, especially example 1) 
\begin{eqnarray}
\psi ( 0^{-}) =s\psi ( 0^{+})
\label{eq:BC}
\end{eqnarray}
where $s=e^{i\theta }$ is a unimodular complex number. See Fig. \ref{fig:jump_delta}.

Specifically, we have the integration by parts formula
\begin{eqnarray*}
\left( \int_{-\infty}^{0^-} + \int_{0^+}^\infty \right) \phi^{\ast }(x) \psi^\prime (x) dx =\\
\phi^{\ast }(0^-) \psi (0^-) - \phi^{\ast }(0^+) \psi (0^+) \\
 -\left( \int_{-\infty}^{0^-} + \int_{0^+}^\infty \right) \phi^{\ast \prime }(x) \psi (x) dx,
\end{eqnarray*}
and the boundary term vanishes exactly if both $\phi$ and $\psi$ satisfy the (same!) boundary condition (\ref{eq:BC})
with $s$ unitary.

As $s$ is arbitrary,
we have an infinity of possible self-adjoint extensions ${H}_{s}$.
Ultimately the choice of $s$ comes down to physical modeling. 

We shall look at two different approximation schemes now.

\subsection{When the phase jump will be an exponential of the coupling parameter.}

We take the potential to have the form 
\begin{eqnarray}
{V}=\delta _{n}( {x})
\end{eqnarray}
We consider a stationary state with far field behaviour $\psi ( x) =A_{\pm
}e^{-ikx}$ for $x \to \pm \infty $. We have $i\psi ^{\prime }(x)+V( x) \psi
( x) =E\psi ( x) $. Away from the origin,\ where the potential is zero, we
see that $E\equiv k$ We can integrate to get 
\begin{eqnarray*}
\psi ( b) e^{ikb}e^{-i\int_{a}^{b}V( x) dx}=\psi ( a) e^{ika}.
\end{eqnarray*}
We have $V( x) =\varepsilon \delta _{n}( x) $, and so taking $n \to \infty$ 
$\psi ( b) e^{ikb}e^{i\varepsilon }=\psi ( a) e^{ika}$ whenever $a<0<b$.
Taking the points $a$ and $b$ to approach the origin yields $\psi ( 0^{-})
=e^{-i\varepsilon }\psi ( 0^{+}) $, that is
\begin{eqnarray}
s= e^{-i\varepsilon }.
\end{eqnarray}

\subsection{When the phase jump will be fractional linear in the coupling parameter.}

We now take the potential to have the form 
\begin{eqnarray}
{V}=|\delta _{n }\rangle \langle \delta _{n }|.
\end{eqnarray}
This leads to the stationary state equation 
\begin{eqnarray*}
i\psi ^{\prime }(x)+\langle \delta
_{n }|\psi \rangle \delta _{n }( x) =k\psi ( x) 
\end{eqnarray*}
and one of the
obvious features is that the function $\delta _{n }$ does not stay in
the Hilbert space as $n \to \infty$. As an ansatz, we try a solution of the
form 
\begin{eqnarray}
\psi _{n }( x) =\alpha _{n }\theta _{n }( x) +\phi ( x)
\end{eqnarray}
where $\alpha _{n }$ is a complex scalar and $\theta _{n }( x)
=\int_{-\infty }^{x}\delta _{n }( x^{\prime }) dx^{\prime }$. The
function $\phi $ is assumed to be continuous and differentiable at $x=0$.
Substituting the trial function, we find 
\begin{eqnarray*}
i\alpha _{n }\delta _{n }( x) +i\phi ^{\prime }( x) +\varepsilon
\langle \delta _{n }|\psi _{n }\rangle \delta _{n }( x)
=k\alpha _{n }\theta _{n }( x) +k\phi ( x) ,
\end{eqnarray*}
and to remove the divergent $\delta _{n }( x) $ term we must take $%
\alpha _{n }=i\varepsilon \langle \delta _{n }|\psi _{n
}\rangle $. This leaves $i\phi ^{\prime }( x) =k\alpha _{n }\theta
_{n }( x) +k\phi ( x) $ which integrates to 
\begin{eqnarray}
e^{ikb}\phi ( b) -e^{ika}\phi ( a) =ik\alpha _{n
}\int_{a}^{b}e^{ikx}\theta _{n }( x) dx
\end{eqnarray}
and the right hand side converges to $\alpha ( e^{ikb}-1) $ as $n \to
\infty $, whenever $a<x<b$. Again, taking $a$ and $b$ approaching zero we get the
consistency condition $\phi ( 0^{-}) =\phi ( 0^{+}) $. The limit function $%
\psi $ is then $\psi ( x) =\alpha \theta ( x) +\phi ( x) $ where $\theta $
is the Heaviside function and $\alpha =\lim_{n \to \infty}\alpha _{n
}=i\varepsilon \frac{\psi ( 0^{+}) +\psi ( 0^{-}) }{2}$. We then have $\psi
( 0^{+}) =\psi ( 0^{-}) +\alpha $. Eliminating $\alpha $ then leads to $\psi
( 0^{-}) =\frac{1-\frac{i}{2}\varepsilon }{1+\frac{i}{2}\varepsilon }\psi (
0^{+}) $, that is \begin{eqnarray}
s= \frac{1-\frac{i}{2}\varepsilon }{1+\frac{i}{2}\varepsilon } .
\end{eqnarray}

\subsection{Remarks}

The choice of $-{p}$ as free Hamiltonian lead to scattering coefficient $%
s$ that is independent of the wavenumber $k$. We encounter two very
different forms: the exponential $s_{1}(\varepsilon )=e^{-i\varepsilon }$
and the fractional linear $s_{2}(\varepsilon )=\frac{1-\frac{i}{2}%
\varepsilon }{1+\frac{i}{2}\varepsilon }$. Remarkably they agree up to
second order when Taylor expanded in the coupling strength $\varepsilon $.
This means that intuitive arguments based on perturbation expansions may not
always be reliable.

\section{Stochastic Jump Evolutions}

We now second quantize the situation encountered in the Section \ref{sec:single}. We
consider a single quantum input process $b( t) $ satisfying singular
commutation relations $[ b( t) ,b( s) ^{\dag }] =\delta ( t-s) $. and set $%
\Lambda ( t) =\int_{0}^{t}b^{\dag }( s) b( s) ds$, also known as the gauge
process. The relevant quantum It\={o} product rule is $( d\Lambda ( t) )
^{2}=d\Lambda ( t) $, and we have the It\={o} formula $df( \Lambda _{t}) =\{
f( \Lambda ( t) +1) -f( \Lambda ( t) ) \} d\Lambda ( t) $. The general form
of a pure-gauge unitary evolution coupling the system to the input field
takes the form \cite{HP} 
\begin{eqnarray}
U( t) =(S\otimes I)^{I\otimes \Lambda ( t) }
\end{eqnarray}
satisfying the quantum stochastic differential equation $dU( t) =( S-1)
\otimes d\Lambda ( t) \,U( t) $, where $S$ is required to be a unitary
operator on the system space. This is a degenerate triple $(
S,0,0) $, that is the coupling parameters $L=0$ and the Hamiltonian $H=0$ in (\ref{eq:SLH}).

We now show the quantum stochastic analogues to the two types of limit encountered in the previous section.
 For convenience, we restrict to a single input
field, but the generalisation to multiple modes is straightforward.

\subsection{When the scattering matrix will be an exponential of the exchange matrix.}

Fix a self-adjoint operator $E_{11}$ on the system Hilbert space, then chose
the time-dependent Hamiltonian 
\begin{eqnarray}
\tilde{H}^{( n) }( t) =E_{11}\otimes \tilde{\lambda}^{( n) }( t) ,
\end{eqnarray}
with $\tilde{\lambda}^{( n) }( t) =\int \delta _{n}( t-s) d\Lambda ( s)
\equiv \int \delta _{n}( t-s) b^{\dag }( s) b( s) ds$. 
We denote by $\tilde{U}^{(n)}(
t) $  the solution to the corresponding Schr\"{o}dinger equation 
\begin{eqnarray}
i\frac{d}{dt}\tilde{U}_{(n)}( t) =\tilde{H}^{( n) }( t) \tilde{ U}^{(n)}( t)
\end{eqnarray}
with $\tilde{U}^{(n)}( 0) =I$. Here the solution will be $\tilde{U}^{(n)}(
t) =\exp \{ -iE_{11}\otimes \int_{0}^{t}ds\int \delta _{n}( s-u) d\Lambda (
u) \} $, but 
\begin{eqnarray}
\int_{0}^{t}ds\int \delta _{n}( s-u) d\Lambda ( u) =\int \delta _{n}\ast
1_{[ 0,t] }( u) d\Lambda ( u)
\end{eqnarray}
and we encounter the convolution $\delta _{n}\ast 1_{[ 0,t] }$ of the
approximate delta function $\delta _{n}$ with the indicator function $1_{[
0,t] }$ of the interval $[ 0,t] $. We then have the strongly convergent limit to $%
\Lambda ( t) $ and so $\tilde{U}^{(n)}( t) $ is strongly convergent to $%
\tilde{U}( t) =\exp \{ -iE_{11}\otimes \Lambda ( t) \} $. This of course
corresponds to the pure gauge driven unitary with scattering matrix $\tilde{S%
}=e^{-iE_{11}}$.

In the multiple input field case we have the matrix relation
\begin{eqnarray}
\tilde{S}=e^{-iE_{\ell \ell}}
\end{eqnarray}
N.B. Recall that the entries $E_{jk}$ of the exchange matrix are operators on the system.

This limit is naturally associated with the Holevo time-ordered exponential form of
the quantum stochastic calculus as, indeed,
\begin{eqnarray}
\tilde{U}(t) =\vec{T}_{H} e^{-i\sum_{jk} E_{jk} \Lambda_{jk}(t)}.
\end{eqnarray}

\subsection{When the scattering matrix will be fractional linear in the exchange matrix.}

We alternatively take 
\begin{eqnarray}
H^{(n)}(t)=E_{11}\otimes \tilde{b}^{(n)}(t)^{\dag }\tilde{b}^{(n)}(t)
\end{eqnarray}
where $\tilde{b}^{(n)}(t)=\int \delta _{n}(t-s)b(s)ds$ is a smeared
annihilator, etc. The unitaries $U_{n}(t)$ generated by time-dependent
Hamiltonian $H^{(n)}(t)$ converge to the unitary quantum stochastic process $U(t)$ 
with triple $(S,0,0)$ where $S=\dfrac{1-\frac{i}{2}E_{11}}{1+\frac{i}{2}E_{11}}$, \cite{WZ}.

In the multiple input field case we then have the matrix relation
\begin{eqnarray}
S= \dfrac{1-\frac{i}{2}E_{\ell \ell}}{1+\frac{i}{2}E_{\ell \ell}}
\end{eqnarray}

This limit is naturally associated with the Stratonovich form as now
\begin{eqnarray}
U(t) =\vec{T}_{D} e^{-i\sum_{jk} E_{jk} \Lambda_{jk}(t)}.
\end{eqnarray}

\subsection{Adiabatic Elimination of a Cavity Mode}

In \cite{AE} the adiabatic elimination of a cavity mode $b$ was considered
where the mode had a Hamiltonian of the form 
\begin{eqnarray*}
H^{\left( n\right) }=E_{00}+\sqrt{n}E_{10}b^{\ast }+\sqrt{n}%
E_{01}b+nE_{11}b^{\dag }b
\end{eqnarray*}
with the mode coupled to an external field with a coupling strength $\sqrt{%
n\gamma }$. That is, we have the QSDE 
\begin{eqnarray*}
dU_{t} = \left\{ \sqrt{n\gamma }b\otimes dB_{t}^{\ast }-\sqrt{n\gamma }b^{\ast
}\otimes dB_{t} \right.  \\
  \left. -\frac{n\gamma }{2}b^{\ast }bdt+i H^{\left( n\right)
}dt\right\} U_{t}.
\end{eqnarray*}
If we now introduce the unperturbed dynamics $dV_{t}=\{ \sqrt{n\gamma }b\otimes
dB_{t}^{\ast }-\sqrt{n\gamma }b^{\ast }\otimes dB_{t}-\frac{n\gamma }{2}%
b^{\ast }bdt\}V_{t}$, then it is shown that the unitary $\tilde{U}%
_{t}=V_{t}^{-1}U_{t}$ satisfies a limit QSDE as $n\rightarrow \infty $ of
the form (\ref{eq:QSDE}) with $\left( S,L,H\right) $ given by 
\begin{eqnarray*}
S &=&\frac{\gamma /2-iE_{11}}{\gamma /2+iE_{11}}, \\
L &=&\frac{i\sqrt{\gamma }}{\gamma /2+iE_{11}}, \\
H &=&E_{00}+E_{01}\text{Im}\left\{ \frac{1}{\gamma /2+iE_{11}}\right\}
E_{01}.
\end{eqnarray*}
In the special case of an atomic system in a cavity, one
may consider \cite{Doherty} 
\begin{eqnarray*}
E_{11}=-\frac{g_{0}^{2}}{\Delta }\cos ^{2}\left( kq\right) ,\text{\ }%
E_{01}=0=E_{10},\;E_{00}=H
\end{eqnarray*}
and so 
\begin{eqnarray}
S &=&\frac{\gamma /2+i\frac{g_{0}^{2}}{\Delta }\cos ^{2}\left( kq\right) }{%
\gamma /2-i\frac{g_{0}^{2}}{\Delta }\cos ^{2}\left( kq\right) }  \notag \\
&=&\exp \left( 2i\tan ^{-1}\left( \frac{2g_{0}^{2}}{\gamma \Delta }\cos
^{2}\left( kq\right) \right) \right) .  \label{eq:S_AE}
\end{eqnarray}
In the limit model, we find that the atomic indeed induces a phase change on the optical field.

\section{Variable-Speed Quantum Traveling Field Modes}

We now give a non-perturbative argument leading to jump QSDEs starting from
scattering of light by quantum systems which correspond to free boundaries.
Our approach is to use the theory developed by Ley and Loudon \cite{Ley+Loudon87}, se also \cite{Blow},
where the quantise the classical mode fields for electomagnetic fields scattered by dielectric media, as
opposed to trying to begin with free photons. We mention other appraoches such as \cite{BBB} which 
deal with quanisation of light in dielectric material.

We begin with Maxwell's equations without sources 
\begin{eqnarray*}
\mathbf{\nabla }.\mathbf{B} &=&0,\mathbf{\nabla }\times \mathbf{E}=-\frac{%
\partial }{\partial t}\mathbf{B}, \\
\mathbf{\nabla }.\mathbf{D} &=&0,\mathbf{\nabla }\times \mathbf{H}=\frac{%
\partial }{\partial t}\mathbf{D}.
\end{eqnarray*}
Our interest is in the situation where the displacement field takes the form 
$\mathbf{D}=\varepsilon \mathbf{E}$ with a dielectric coefficient $%
\varepsilon $ which depends on position. We shall take $\mathbf{B}=\mu _{0}%
\mathbf{H}$ with constant permeability $\mu _{0}$. The first pair of
equations lead to the usual potential 
\begin{eqnarray*}
\mathbf{E}=-\frac{\partial }{\partial t}\mathbf{A}-\mathbf{\nabla }\phi
,\quad \mathbf{B}=\mathbf{\nabla }\times \mathbf{A},
\end{eqnarray*}
and we will fix the Coulomb gauge $\left( \phi \equiv 0\right) $.

We seek to model a field propagating in a thin wire along the $z$-axis of
cross-section $\mathcal{A}$, and to this end we take the vector potential to
have non-zero component $A_{x}=A_{x}\left( z,t\right) $ in which case the
non-zero components of the electric and magnetic fields are 
\begin{eqnarray*}
E_{x}\left( z,t\right) =-\frac{\partial }{\partial t}A_{x}\left( z,t\right)
,\quad B_{y}\left( z,t\right) =\frac{\partial }{\partial z}A_{x}\left(
z,t\right) .
\end{eqnarray*}
We also take the dielectric constant to be a function of the $z$ coordinate
only and set 
\begin{eqnarray}
\zeta \left( z\right) =\varepsilon \left( z\right) \mu _{0}\equiv \frac{1}{%
c^{2}}n\left( z\right) ^{2}  \label{eq: zeta = n2 /c2}
\end{eqnarray}
where $n\left( z\right) $ is a position dependent refractive index. We shall
assume the asymptotic behaviour 
\begin{eqnarray*}
\lim_{z\rightarrow \infty }\zeta \left( z\right) =\left( \frac{n_{r}}{c}%
\right) ^{2},\,\lim_{z\rightarrow -\infty }\zeta \left( z\right) =\left( 
\frac{n_{l}}{c}\right) ^{2}.
\end{eqnarray*}
Here $n_{r}\geq 1$ and $n_{l}\geq 1$ are the refractive indices in the right
and left far zones respectively. The equation $\mathbf{\nabla }.\mathbf{D}=0$
is now trivially satisfied, and with the remaining Maxwell's equation we see
that the component $A_{x}$ satisfies the wave equation 
\begin{eqnarray}
\left( \frac{\partial ^{2}}{\partial z^{2}}-\zeta \left( z\right) \frac{%
\partial ^{2}}{\partial t^{2}}\right) A_{x}\left( z,t\right) =0.
\label{eq:diff-A}
\end{eqnarray}

\subsection{Mode Functions}

Let $\omega >0$ be a positive frequency, and consider trial solutions to (\ref{eq:diff-A}) 
of the form $U_{\omega }\left( z\right) e^{-i\omega t}$.
We see that the mode functions $U_{z}$ satisfy 
\begin{eqnarray}
\left( \frac{d^{2}}{dz^{2}}+\zeta \left( z\right) \omega ^{2}\right)
U_{\omega }\left( z\right) =0.  \label{eq:diff-U}
\end{eqnarray}
The mode corresponding to a right incoming traveling wave is the solution $%
U_{\omega ,r}$ with the asymptotic behavior 
\begin{eqnarray*}
U_{\omega ,r}\left( z\right) \simeq \left\{ 
\begin{array}{cc}
e^{-in_{r}\omega z/c}+\mathsf{t}_{rr}\left( \omega \right) e^{in_{r}\omega
z/c}, & z\rightarrow \infty ; \\ 
\mathsf{t}_{lr}\left( \omega \right) e^{-in_{l}\omega z/c}, & z\rightarrow
-\infty .
\end{array}
\right.
\end{eqnarray*}
We may similarly introduce the left incoming mode as the solution with the
asymptotic behavior 
\begin{eqnarray*}
U_{\omega ,l}\left( z\right) \simeq \left\{ 
\begin{array}{cc}
\mathsf{t}_{rl}\left( \omega \right) e^{in_{r}\omega z/c}, & z\rightarrow
\infty ; \\ 
e^{in_{l}\omega z/c}+\mathsf{t}_{ll}\left( \omega \right) e^{-in_{l}\omega
z/c}, & z\rightarrow -\infty .
\end{array}
\right.
\end{eqnarray*}

\begin{figure}[htbp]
\centering
\includegraphics[width=0.450\textwidth]{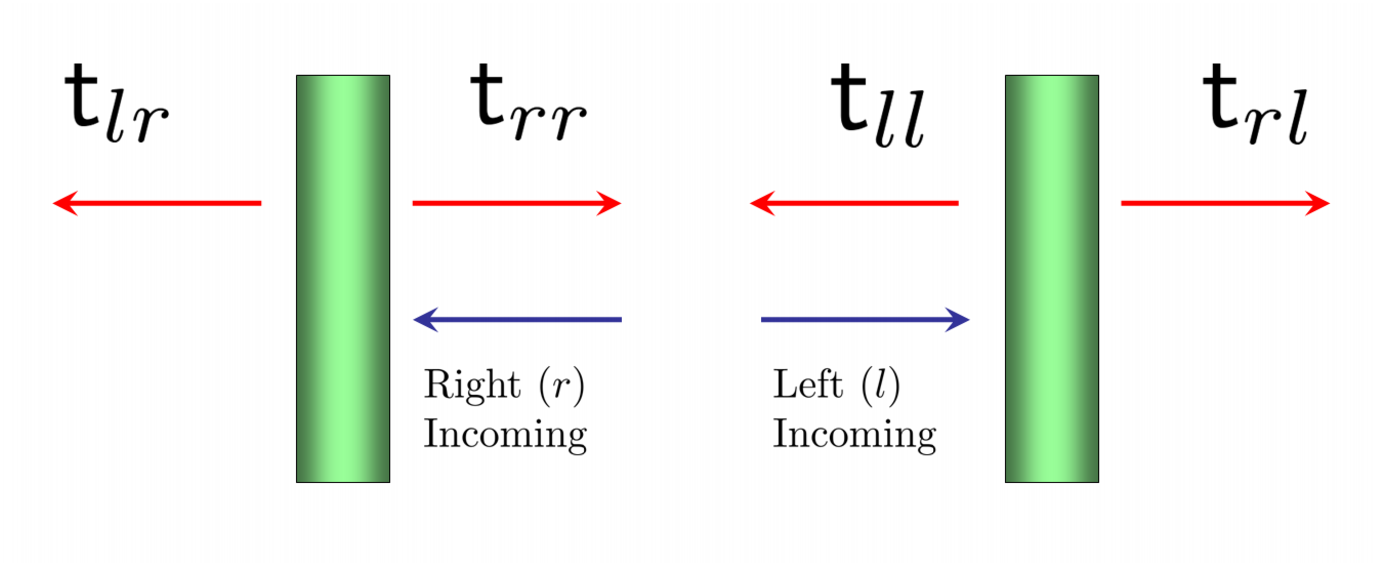}
\caption{(Color online) In the far zone: the right incoming plane wave has reflected
coefficient $\mathsf{t}_{rr}$ and transmitted coefficient $\mathsf{t}_{lr}$; likewise the left
incoming plane wave has reflected coefficient $\mathsf{t}_{ll}$ and transmitted
coefficient $\mathsf{t}_{rl}$.}
\label{fig:Jump_tc}
\end{figure}

See Fig. \ref{fig:Jump_tc}.

The fields may be decomposed as $A\left( z,t\right) =A^{+}\left( z,t\right)
+A^{-}\left( z,t\right) $ where positive-frequency components are then given
by 
\begin{eqnarray*}
A^{+}\left( z,t\right) &=&\int_{0}^{\infty }\sqrt{\frac{\hbar }{2\pi 
\mathcal{A}\omega }}\sum_{c=r,l}U_{\omega ,c}\left( z\right) a_{c}\left(
\omega \right) e^{-i\omega t}d\omega \\
E^{+}\left( z,t\right) &=&i\int_{0}^{\infty }\sqrt{\frac{\hbar \omega }{2\pi 
\mathcal{A}}}\sum_{c=r,l}U_{\omega ,c}\left( z\right) a_{c}\left( \omega
\right) e^{-i\omega t}d\omega \\
B^{+}\left( z,t\right) &=&\int_{0}^{\infty }\sqrt{\frac{\hbar }{2\pi 
\mathcal{A}\omega }}\sum_{c=r,l}U_{\omega ,c}^{\prime }\left( z\right)
a_{c}\left( \omega \right) e^{-i\omega t}d\omega
\end{eqnarray*}
while the negative-frequency fields are just the hermitean conjugates $A^{-}\left( z,t\right)
=A^{+}\left( z,t\right) ^{\dag }$, etc. The field is quantized by
introducing the canonical commutation relations 
\begin{eqnarray}
\left[ a_{b}\left( \omega \right) ,a_{c}^{\dag }\left( \omega ^{\prime
}\right) \right] =\delta _{bc}\,\delta \left( \omega -\omega ^{\prime
}\right) ,  \label{eq: CCR}
\end{eqnarray}
where $b,c\in \left\{ r,l\right\} $.

\subsection{Far Field Input/Output Relations}

The electromagnetic field has the (Abrahams) momentum density is $\mathbf{g}=%
\frac{1}{c^{2}}\mathbf{E}\times \mathbf{H}$ which in the present case has
non-zero component $g_{z}=\frac{1}{\varepsilon _{0}}E_{x}B_{y}$. Taking a 
\emph{classical} potential $A^{+}\left( z,t\right) =\sum_{c=r,l}U_{\omega
,c}\left( z\right) \alpha _{c}e^{-i\omega t}$ for a pair of complex
constants $\alpha _{r},\alpha _{l}$. The far right and far left one-cycle
time-averaged values should be equal in order to have conservation of
electromagnetic momentum and from this we derived the following identities 
\begin{eqnarray*}
n_{r}\left| \mathsf{t}_{rr}\left( \omega \right) \right| ^{2}+n_{l}\left| 
\mathsf{t}_{lr}\left( \omega \right) \right| ^{2} &=&n_{r}, \\
n_{r}\left| \mathsf{t}_{rl}\left( \omega \right) \right| ^{2}+n_{l}\left| 
\mathsf{t}_{ll}\left( \omega \right) \right| ^{2} &=&n_{l}, \\
n_{r}\mathsf{t}_{rr}\left( \omega \right) ^{\ast }\mathsf{t}_{rl}\left(
\omega \right) +n_{l}\mathsf{t}_{lr}\left( \omega \right) ^{\ast }\mathsf{t}%
_{ll}\left( \omega \right) &=&0,
\end{eqnarray*}
due to the arbitrariness of $\alpha _{r}$ and $\alpha _{l}$. A little
algebra also reveals that we have $n_{l}^{2}\left| \mathsf{t}_{lr}\left(
\omega \right) \right| ^{2}=n_{r}^{2}\left| \mathsf{t}_{rl}\left( \omega
\right) \right| ^{2}$. The matrix $\mathsf{t}\left( \omega \right) =\left[ 
\begin{array}{cc}
\mathsf{t}_{rr}(\omega ) & \mathsf{t}_{rl}(\omega ) \\ 
\mathsf{t}_{lr}(\omega ) & \mathsf{t}_{ll}(\omega )
\end{array}
\right] $ will be unitary in the special case where the left and right
refractive indices are equal ($n_{l}=n_{r}$).

In general however the relations between the left and right far field
components can expressed by saying that the matrix 
\begin{eqnarray}
S(\omega )=\left[ 
\begin{array}{cc}
\mathsf{t}_{rr}(\omega ) & \sqrt{\dfrac{n_{r}}{n_{l}}}\mathsf{t}_{rl}(\omega
) \\ 
\sqrt{\dfrac{n_{l}}{n_{r}}}\mathsf{t}_{lr}(\omega ) & \mathsf{t}_{ll}(\omega
)
\end{array}
\right]  \label{eq:S-matrix from T}
\end{eqnarray}
is unitary.

\subsection{Orthogonality of the Modes}

We now generalize the results of Ley and Loudon \cite{Ley+Loudon87} to $%
n_{r},n_{l}\neq 1$. From the observation that 
\[ \left( U_{\omega _{1}}^{\ast
}U_{\omega _{2}}^{\prime }-U_{\omega _{1}}^{\ast \prime }U_{\omega
_{2}}\right) ^{\prime }=\left( \omega _{1}^{2}-\omega _{2}^{2}\right) \zeta
U_{\omega _{1}}^{\ast }U_{\omega _{2}}
\]
we see that 
\begin{eqnarray*}
&&\int_{-M}^{L}U_{\omega _{1}}^{\ast }\left( z\right) U_{\omega _{2}}\left(
z\right) \zeta \left( z\right) dz \\
&=&\frac{1}{\left( \omega _{1}^{2}-\omega _{2}^{2}\right) }\left. \left(
U_{\omega _{1}}^{\ast }U_{\omega _{2}}^{\prime }-U_{\omega _{1}}^{\ast
\prime }U_{\omega _{2}}\right) \right| _{z=-M}^{L}.
\end{eqnarray*}

For the right incoming mode we have, for large $M$ and $L$, 
\begin{eqnarray*}
&&\frac{1}{\left( \omega _{1}^{2}-\omega _{2}^{2}\right) }\left\{
e^{in_{r}\left( \omega _{1}-\omega _{2}\right) L/c}\frac{\left( \omega
_{1}+\omega _{2}\right) n_{r}}{ic}\right. \\
&&+e^{-in_{r}\left( \omega _{1}-\omega _{2}\right) L/c}\frac{\left( \omega
_{1}+\omega _{2}\right) n_{r}}{ic}\mathsf{t}_{rr}\left( \omega _{1}\right)
^{\ast }\mathsf{t}_{rr}\left( \omega _{2}\right) \\
&&\left. +e^{-in_{l}\left( \omega _{1}-\omega _{2}\right) M/c}\frac{\left(
\omega _{1}+\omega _{2}\right) n_{l}}{ic}\mathsf{t}_{lr}\left( \omega
_{1}\right) ^{\ast }\mathsf{t}_{lr}\left( \omega _{2}\right) +\cdots
\right\} .
\end{eqnarray*}
The remaining terms, appearing as an ellipsis, are proportional to $\left( \omega _{1}-\omega
_{2}\right) $ $\times e^{\pm i\left( n_{r}\omega _{1}L+n_{l}\omega
_{2}M\right) /c}$ and will not contribute. We need to take $M=\frac{n_{r}}{%
n_{l}}L$ to obtain a balanced limit which leads us to define the following
skewed principal value integral: 
\begin{eqnarray*}
-\hspace{-0.15in}\int_{-\infty }^{\infty }f\left( z\right)
dz:=\lim_{N\rightarrow \infty }\int_{-N/n_{l}}^{N/n_{r}}f\left( z\right) dz
\end{eqnarray*}
in which case 
\begin{eqnarray*}
&&-\hspace{-0.15in}\int_{-\infty }^{\infty }U_{\omega _{1},r}^{\ast }\left(
z\right) U_{\omega _{2},r}\left( z\right) \zeta \left( z\right) dz \\
&=&\lim_{N\rightarrow \infty }2N\,\text{sinc}\left( N\left( \omega
_{1}-\omega _{2}\right) \right) \\
&=&2\pi \delta \left( \omega _{1}-\omega _{2}\right) ,
\end{eqnarray*}
where sinc$\left( x\right) =\frac{\sin x}{x}$. Here we use the flux
relations with $\omega \rightarrow 0$. The first three terms contribute
while the remainder contributes an unsupported $\delta \left( \omega
_{1}+\omega _{2}\right) $ which is ignored. (Note that the sinc function is
an improper integral and not absolutely integrable, therefore the choice of
skew adopted is necessary to obtain the nascent delta function limit.)

One finds that the various modes are orthogonal in the sense that 
\begin{eqnarray*}
\int_{-\infty }^{\infty }U_{\omega _{1},b}^{\ast }\left(
z\right) U_{\omega _{2},c}\left( z\right) \zeta \left( z\right) dz=2\pi
\delta _{bc}\delta \left( \omega _{1}-\omega _{2}\right) .
\label{q: orthog - U}
\end{eqnarray*}
The sesquilinear form appearing here is the correct notion to formulate the
Sturm-Liouville orthogonality property for the modes.

\subsection{The Hamiltonian}

The Hamiltonian is then taken to be 
\begin{eqnarray}
H &=&\frac{1}{2}-\hspace{-0.15in}\int \left[ \varepsilon \left( z\right)
E^{-}E^{+}\left( z,t\right) +\frac{1}{\mu _{0}}B^{-}B^{+}\left( z,t\right) %
\right] dV  \notag \\
&=&\int_{0}^{\infty }\hbar \omega \left[ a_{r}^{\dag }\left( \omega \right)
a_{r}\left( \omega \right) +a_{l}^{\dag }\left( \omega \right) a_{l}\left(
\omega \right) \right] d\omega ,  \label{eq: H}
\end{eqnarray}
where the $z$-integration is interpreted as the skewed principal value.

\section{Two-Sided Models}

In this section we present several situations of interest.

\subsection{Left and Right regions with unequal constant dielectric
coefficient}

We have a constant dielectric coefficients in two semi-infinite regions with
boundary at $z=q$: 
\begin{eqnarray*}
\zeta \left( z\right) =\left\{ 
\begin{array}{cc}
\left( \frac{n_{r}}{c}\right) ^{2}, & z>q; \\ 
\left( \frac{n_{l}}{c}\right) ^{2}, & z<q.
\end{array}
\right.
\end{eqnarray*}
The modes will be piecewise plane waves, for instance, 
\begin{eqnarray*}
U_{\omega ,r}\left( z\right) =\left\{ 
\begin{array}{cc}
e^{-in_{r}\omega z/c}+\mathsf{t}_{rr}\left( \omega \right) e^{in_{r}\omega
z/c}, & z>q; \\ 
\mathsf{t}_{lr}\left( \omega \right) e^{-in_{r}\omega z/c}, & z<q.
\end{array}
\right.
\end{eqnarray*}
Continuity of $U_{\omega ,r}$ and $U_{\omega ,r}^{\prime }$ across the
boundary yields 
\begin{eqnarray*}
\mathsf{t}_{rr}\left( \omega \right) &=&-\frac{n_{l}-n_{r}}{n_{l}+n_{r}}%
e^{-2in_{r}\omega q/c}, \\
\mathsf{t}_{lr}\left( \omega \right) &=&\frac{2n_{r}}{n_{l}+n_{r}}%
e^{-i(n_{r}-n_{l})\omega q/c}.
\end{eqnarray*}
and one similarly calculates that $\mathsf{t}_{ll}\left( \omega \right) =%
\frac{n_{l}-n_{r}}{n_{l}+n_{r}}e^{2in_{l}\omega q/c}$, $\mathsf{t}%
_{rl}\left( \omega \right) =\frac{2n_{l}}{n_{l}+n_{r}}e^{-i(n_{r}-n_{l})%
\omega q/c}$. The matrix $S\left( \omega \right) $ is then given by 

\begin{eqnarray}
\left[ 
\begin{array}{cc}
\frac{\left( n_{r}-n_{l}\right)}{n_{r}+n_{l}} e^{-2in_{r}\omega q/c} & \frac{2\sqrt{n_{r}n_{l}}}{n_{r}+n_{l}}%
e^{-i(n_{r}-n_{l})\omega q/c} \\ 
\frac{2\sqrt{n_{r}n_{l}}}{n_{r}+n_{l}}e^{-i(n_{r}-n_{l})\omega q/c} & \frac{\left( n_{l}-n_{r}\right)}{n_{r}+n_{l}}
e^{2in_{l}\omega q/c}
\end{array}
\right] . 
 \label{eq: S_boundary}
\end{eqnarray}

\subsection{Dielectric Slab}

A dielectric slab of thickness $2a$ about $z=q$ is modeled by 
\begin{eqnarray*}
c^{2}\zeta \left( z\right) =\left\{ 
\begin{array}{ll}
n_{r}^{2}, & I:z>q+a; \\ 
n^{2}, & II:|z-q|<a; \\ 
n_{l}^{2}, & III:z<q-a.
\end{array}
\right.
\end{eqnarray*}
We now set 
\begin{eqnarray*}
U_{\omega ,r}\left( z\right) =\left\{ 
\begin{array}{ll}
e^{-in_{r}\omega z/c}+\mathsf{t}_{rr}\left( \omega \right) e^{in_{r}\omega
z/c}, & I; \\ 
A\left( \omega \right) e^{-in\omega z/c}+B\left( \omega \right) e^{in\omega
z/c}, & II; \\ 
\mathsf{t}_{lr}\left( \omega \right) e^{-in_{l}\omega z/c}, & III.
\end{array}
\right.
\end{eqnarray*}
Again requiring continuity of $U_{\omega ,r}$ and $U_{\omega ,r}^{\prime }$
across the boundaries leads to the expressions for $\mathsf{t}_{rr}\left(
\omega \right) $ and $\mathsf{t}_{lr}\left( \omega \right) $: 
\begin{eqnarray*}
\mathsf{t}_{rr}\left( \omega \right) &=&\frac{1}{D\left( \omega \right) }%
\{\left( n+n_{l}\right) \left( n-n_{r}\right) e^{2ina\omega /c} \\
&&-\left( n-n_{l}\right) \left( n+n_{r}\right) e^{-2ina\omega
/c}\}e^{-2in_{r}\left( q+a\right) \omega /c}, \\
\mathsf{t}_{lr}\left( \omega \right) &=&-\frac{1}{D\left( \omega \right) }%
4nn_{r}e^{i(n_{l}-n_{r})q\omega /c}e^{-i\left( n_{r}+n_{l}\right) a\omega
/c},
\end{eqnarray*}
with the denominator $D\left( \omega \right) $%
\begin{eqnarray*}
\left( n-n_{r}\right) \left( n-n_{l}\right) e^{2ina\omega /c}-\left(
n+n_{r}\right) \left( n+n_{l}\right) e^{-2ina\omega /c}.
\end{eqnarray*}

The left incoming coefficients are obtained by symmetry: $\mathsf{t}%
_{rr}\hookrightarrow \mathsf{t}_{ll}$ and $\mathsf{t}_{rl}\hookrightarrow 
\mathsf{t}_{lr}$ under the parameter inversion $a\hookrightarrow -a$, $%
n_{r}\hookrightarrow -n_{l}$ and $n_{l}\hookrightarrow -n_{r}$.

\subsection{Singular Dielectric Boundaries}

It is of interest to consider the limit of vanishing thickness with large
internal refractive index. Specifically we consider the previous model of a
dielectric slab and take the limits 
\begin{eqnarray*}
a\rightarrow 0\text{ with }2n^{2}a=\mu \text{ (constant).}
\end{eqnarray*}
The limiting forms are then

\begin{eqnarray*}
\mathsf{t}_{rr}\left( \omega \right) &=&\frac{\left( n_{r}-n_{l}\right)
+i\mu \omega /c}{\left( n_{r}+n_{l}\right) -i\mu \omega /c}e^{-2in_{r}\omega
q/c}, \\
\mathsf{t}_{lr}\left( \omega \right) &=&\frac{2n_{r}}{\left(
n_{r}+n_{l}\right) -i\mu \omega /c}e^{i\left( n_{l}-n_{r}\right) \omega q/c}.
\end{eqnarray*}

The left incoming coefficients are similarly calculated (the parameter
inversion is now $\mu \hookrightarrow -\mu$, $n_{r}\hookrightarrow -n_{l}$
and $n_{l}\hookrightarrow -n_{r}$) and one has 
\begin{multline*}
S\left( \omega \right) =\frac{1}{\left( n_{r}+n_{l}\right) -i\mu \omega /c}%
\times \\
\left[ 
\begin{array}{cc}
\left( n_{r}-n_{l}+i\dfrac{\mu \omega }{c}\right) e^{-2in_{r}\frac{\omega }{c%
}q} & 2\sqrt{n_{r}n_{l}}e^{i\left( n_{l}-n_{r}\right) \frac{\omega }{c}q} \\ 
2\sqrt{n_{r}n_{l}}e^{i\left( n_{l}-n_{r}\right) \frac{\omega }{c}q} & \left(
n_{l}-n_{r}+i\dfrac{\mu \omega }{c}\right) e^{2in_{l}\frac{\omega }{c}q}
\end{array}
\right]
\end{multline*}

\subsection{Singular Dielectric Points}

In particular, if we set $n_{l}=n_{r}=1$, then we are left with the singular
dielectric point at $z=q$: 
\begin{eqnarray}
S\left( \omega \right) =\frac{1}{2-i\mu \omega /c}\left[ 
\begin{array}{cc}
\dfrac{i\mu \omega }{c}e^{-2i\omega q/c} & 2 \\ 
2 & -\dfrac{i\mu \omega }{c}e^{2i\omega q/c}
\end{array}
\right] .  \label{eq:S_singular_point}
\end{eqnarray}

This may be formally understood as arising from the singular distribution 
\begin{eqnarray*}
\zeta \left( z\right) =1+\mu \delta \left( z-q\right) .
\end{eqnarray*}
To see this, we take the general solution for $U_{\omega ,r}$, 
\begin{eqnarray*}
U_{\omega ,r}\left( z\right) =\left\{ 
\begin{array}{cc}
e^{-i\omega z/c}+\mathsf{t}_{rr}\left( \omega \right) e^{i\omega z/c}, & z>q;
\\ 
\mathsf{t}_{lr}\left( \omega \right) e^{-i\omega z/c}, & z<q;
\end{array}
\right.
\end{eqnarray*}
and impose continuity of $U_{\omega ,r}$ at the boundary $q$, along with the
condition 
\begin{eqnarray}
U_{\omega ,r}^{\prime }\left( q^{-}\right) -U_{\omega ,r}^{\prime }\left(
q^{+}\right) =\mu \frac{\omega ^{2}}{c^{2}}U_{\omega ,r}\left( q\right) .
\label{eq: BC_jump}
\end{eqnarray}
Physically this is interpreted as a discontinuity in $B_{y}$ due the finite
change in the time-derivative of the displacement $D_{x}$ across the
infinitesimal boundary \cite{Ley+Loudon87}. A similar condition applies to $%
U_{\omega ,l}$.

\section{Mirrors}

Mirrors are special cases where the light comes exclusively from one
direction (the right say) and is reflected back. This leads to a
semi-infinite geometry.

\subsection{Perfect Mirrors}

Perfect Mirrors can be viewed as the limiting situation where, say, the left
refractive index $n_{l}$ becomes infinite. For instance, taking a boundary
at $z=q$, we ignore the left-incoming wave and find that the right incoming
wave is reflected with unimodular coefficient 
\begin{eqnarray}
\mathsf{r}_{r}\left( \omega \right) =\lim_{n_{l}\rightarrow \infty }\mathsf{t%
}_{rr}\left( \omega \right) =-e^{-2in_{r}\omega q/c},
\label{eq:perfect_mirror}
\end{eqnarray}
where we take the $n_{l}\rightarrow \infty $ limit of (\ref{eq: S_boundary}%
). Here the phase includes the information of the boundary position $q$.

\subsection{Singular Dielectric Boundary}

An alternative model would be to have a fixed perfect mirror at $z=0$ and a
singular dielectric boundary at $z=q$. In this case we have $z\geq 0$ only
and set 
\begin{eqnarray*}
U_{\omega ,r}\left( z\right) =\left\{ 
\begin{array}{cc}
e^{-i\omega z/c}+\mathsf{r}_{r}\left( \omega \right) e^{i\omega z/c}, & z>q;
\\ 
A\left( \omega \right) e^{-i\omega z/c}+B\left( \omega \right) e^{i\omega
z/c}, & 0\leq z<q.
\end{array}
\right.
\end{eqnarray*}
The boundary conditions are $U_{\omega ,r}\left( 0\right) =0$ (so $A\left(
\omega \right) =-B\left( \omega \right) $), $U_{\omega ,r}\left(
q^{-}\right) =U_{\omega ,r}\left( q^{+}\right) $, and (\ref{eq: BC_jump}).
One finds that $\mathsf{r}_{r}\left( \omega \right) $ is again unimodular
and given by 
\begin{eqnarray}
\mathsf{r}_{r}\left( \omega \right) =\frac{1-i\frac{\mu \omega }{2c}\left(
e^{-2i\omega q/c}-1\right) }{1+i\frac{\mu \omega }{c}\left( e^{2i\omega
q/c}-1\right) }.  \label{eq:R_sdb}
\end{eqnarray}
We note that $\mathsf{r}_{r}\left( \omega \right) \rightarrow -e^{-2i\omega
q/c}$ as the strength $\mu $ of the infinitesimally thin dielectric becomes
infinite: that is we recover the limit of a perfect mirror at $z=q$.

\subsection{Dielectric Layer}

\begin{figure}[tbph]
\centering
\includegraphics[width=0.40\textwidth]{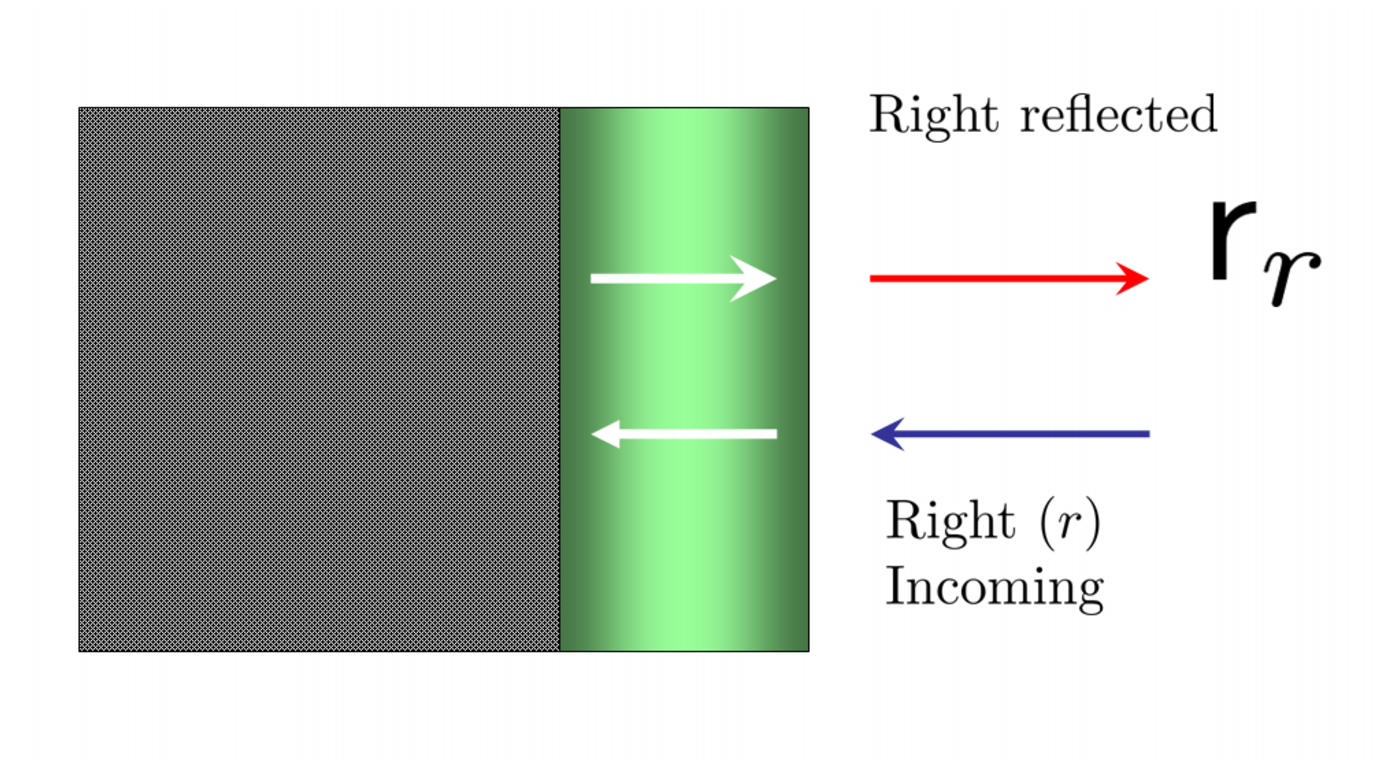}
\caption{(Color online) An imperfect mirror may be modelled as a perfect mirror with a dielectric
layer of refractive index $n$.}
\label{fig:Jump_layer}
\end{figure}

A similar situation is to have a perfect mirror at $z=0$ and a dielectric
slab between 0 and $z=q$ with fixed refractive index $n$, see Fig. \ref{fig:Jump_layer}. It is not too
difficult to show that the reflection is now given by 
\begin{eqnarray*}
\mathsf{r}_{r}\left( \omega \right) =-e^{-2i\omega q/c}\frac{\cos \left( 
\frac{n\omega q}{c}\right) +\frac{i}{n}\sin \left( \frac{n\omega q}{c}%
\right) }{\cos \left( \frac{n\omega q}{c}\right) -\frac{i}{n}\sin \left( 
\frac{n\omega q}{c}\right) }.
\end{eqnarray*}

\section{Input-Output Formulation}

We now wish to introduce an input-output formalism based on measurements by
detectors at positions $Z_{R}$ and $Z_{L}$ positioned in the right and left
far zones of the field respectively. To avoid an number of issues involved
with the one dimensional nature of the fields, we shall assume that both far
zones are non-dielectric, that is 
\begin{eqnarray*}
n_{r}=n_{l}=1.
\end{eqnarray*}
Note that we now have 
\begin{eqnarray}
S(\omega )\equiv \left[ 
\begin{array}{cc}
\mathsf{t}_{rr}(\omega ) & \mathsf{t}_{rl}(\omega ) \\ 
\mathsf{t}_{lr}(\omega ) & \mathsf{t}_{ll}(\omega )
\end{array}
\right]
\end{eqnarray}
which will be unitary.

The electric field at the right detector is given approximately by 
\begin{eqnarray*}
E^{+}\left( Z_{R},t\right) &\simeq& i\int_{0}^{\infty }\sqrt{\frac{\hbar
\omega }{2\pi \mathcal{A}}}e^{-i\omega (t+Z_{R}/c)}a_{r}(\omega )d\omega \\
&+&i\int_{0}^{\infty }\sqrt{\frac{\hbar \omega }{2\pi \mathcal{A}}}\mathsf{t}%
_{rr}\left( \omega \right) e^{-i\omega (t-Z_{R}/c)}a_{r}(\omega )d\omega \\
&+&i\int_{0}^{\infty }\sqrt{\frac{\hbar \omega }{2\pi \mathcal{A}}}\mathsf{t}%
_{rl}\left( \omega \right) e^{-i\omega (t-Z_{R}/c)}a_{l}(\omega )d\omega .
\end{eqnarray*}
We now make the standard quantum white noise assumption for optical fields:
this amounts to identifying a central frequency $\Omega $ and replace the $%
\sqrt{\omega }$ and \textsf{t}$_{ab}\left( \omega \right) $ terms with their
values at $\omega =\Omega $, and otherwise taking the lower value of the
integral to $-\infty $: 
\begin{eqnarray*}
E^{+}\left( Z_{R},t\right) &\simeq &i\sqrt{\frac{\hbar \Omega }{2\pi 
\mathcal{A}}}b_{r}\left( t+Z_{R}/c\right) \\
&&+i\sqrt{\frac{\hbar \Omega }{2\pi \mathcal{A}}}\mathsf{t}_{rr}\left(
\Omega \right) b_{r}\left( t-Z_{R}/c\right) \\
&&+i\sqrt{\frac{\hbar \Omega }{2\pi \mathcal{A}}}\mathsf{t}_{rl}\left(
\Omega \right) b_{l}\left( t-Z_{R}/c\right) .
\end{eqnarray*}
where $b_{k}\left( \tau \right) =\int_{-\infty }^{\infty }e^{-i\omega \tau
}a_{k}(\omega )d\omega $ for $k=r,l$.

Measurement of the electric field may then effectively is a measurement
(e.g., homodyne quadrature, photon counting, etc.) of the scattered output
field 
\begin{eqnarray*}
b_{r}^{\text{out}}\left( t\right) =\mathsf{t}_{rr}\left( \Omega \right)
b_{r}\left( t-Z_{R}/c\right) +\mathsf{t}_{rl}\left( \Omega \right)
b_{l}\left( t-Z_{R}/c\right)
\end{eqnarray*}
with a similar expression for the left field. Ignoring the time delays, we
may write the input-output equations as 
\begin{eqnarray*}
\left[ 
\begin{array}{c}
b_{r}^{\text{out}}\left( t\right) \\ 
b_{l}^{\text{in}}\left( t\right)
\end{array}
\right] =\left[ 
\begin{array}{cc}
S_{rr} & S_{rl} \\ 
S_{lr} & S_{ll}
\end{array}
\right] \left[ 
\begin{array}{c}
b_{r}\left( t\right) \\ 
b_{l}\left( t\right)
\end{array}
\right] .
\end{eqnarray*}

The coefficients $S_{ab}$ are the transmission/reflection coefficients $%
\mathsf{t}_{ab}\left( \Omega \right) $ and it is convenient to introduce the
photon momentum 
\begin{eqnarray*}
k=\Omega /c.
\end{eqnarray*}
We also have that these coefficients depend on the details of (one or more)
free boundaries $q$. We now lift the condition that the $q$ need to be fixed
parameters and allow them to be quantum mechanical. In particular, they are
time varying also.

\subsection{QSDE Model}

We now introduce a quantum stochastic differential equation (QSDE) that
leads to the above input-output models. We consider the QSDE corresponding
to vacuum inputs: 
\begin{eqnarray*}
\dot{U}_{t}=b_{a}\left( t\right) ^{\ast }S_{ab}U_{t}b_{b}\left( t\right)
-iHU_{t}
\end{eqnarray*}
with initial condition $U_{0}=1$. (Here repeated indices imply a sum over
the values $r$ and $l$.) The Hamiltonian term is taken to have a standard
form 
\begin{eqnarray*}
H=\frac{1}{2m}p^{2}+V\left( q\right) .
\end{eqnarray*}
Let us define the processes 
\begin{eqnarray*}
B_{a}\left( t\right) =\int_{0}^{t}b_{a}(t^{\prime })dt^{\prime },\quad
\Lambda _{ab}\left( t\right) =\int_{0}^{t}b_{a}\left( t^{\prime }\right)
^{\ast }b_{b}\left( t^{\prime }\right) dt^{\prime }
\end{eqnarray*}
then the QSDE may be written alternatively as 
\begin{eqnarray*}
dU_{t}=\left\{ \left( S_{ab}-\delta _{ab}\right) d\Lambda _{ab}-iHdt\right\}
U_{t}.
\end{eqnarray*}
The coefficients $S_{ab}$ are now taken to depend on the position operator $%
q $ of the free quantum boundary. They make up a scattering matrix $S$ which
is therefore a two-by-two matrix with position-operator dependent entries,
and $S$ is unitary.

\begin{figure}[htbp]
\centering
\includegraphics[width=0.450\textwidth]{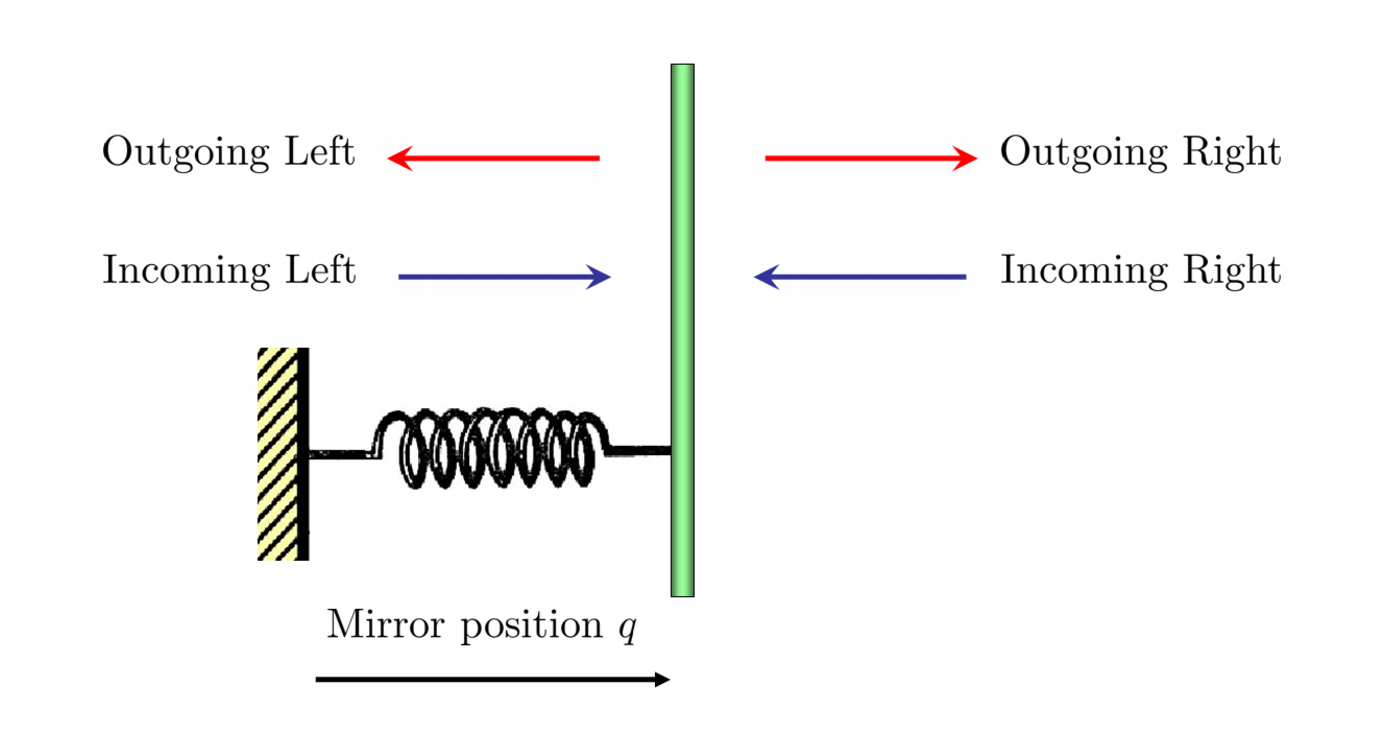} \label{fig:Jump_dielectron}
\caption{(Color online) Dielectric \lq\lq particle\rq\rq scattered by light.}
\end{figure}

Given an arbitrary observable $X$ of the system, its value at time $t$ will
be 
\begin{eqnarray*}
X_{t}=U_{t}^{\ast }\left( X\otimes I\right) U_{t}
\end{eqnarray*}
and from the quantum It\={o} calculus we find 
\begin{eqnarray*}
\dot{X}_{t}=b_{a}^{\ast }\left( t\right) \left( \mathcal{L}_{ab}X\right)
_{t}b_{t}-i\left[ X,H\right] _{t}
\end{eqnarray*}
where 
\begin{eqnarray*}
\mathcal{L}_{ab}X=S_{ca}^{\ast }XS_{cb}-\delta _{ab}X.
\end{eqnarray*}
Now as the $S_{ab}$ are functions of the position observable $q$ only we
have 
\begin{eqnarray*}
\mathcal{L}_{ab}q &=&0, \\
\mathcal{L}_{ab}p &=&-i\hbar S_{ca}^{\ast }S_{cb}^{\prime }
\end{eqnarray*}
where the prime denotes differentiation with respect to the \textit{variable}
$q$. We therefore obtain the position and momentum QSDEs 
\begin{eqnarray}
\dot{q}_{t} &=&\frac{1}{m}p_{t},  \notag \\
\dot{p}_{t} &=&-V^{\prime }\left( q_{t}\right) -i\hbar b_{a}^{\ast }\left(
t\right) \left( S_{ca}^{\ast }S_{cb}^{\prime }\right) b_{b}\left( t\right) .
\label{eq:QSDE_S}
\end{eqnarray}

The vacuum average yields the usual Ehrenfest equations, however, to obtain
something nontrivial we consider a coherent state input field with intensity 
$\beta _{a}(t)$, for $a=r,l$. This is equivalent to making the replacement $%
b_{a}\left( t\right) \rightarrow b_{a}(t)+\beta _{a}(t)$ so that the QSDE
becomes 
\begin{eqnarray*}
\dot{U}_{t}=\left( b_{a}\left( t\right) +\beta _{a}\left( t\right) \right)
^{\ast }S_{ab}U_{t}\left( b_{b}\left( t\right) +\beta _{b}\left( t\right)
\right) ,
\end{eqnarray*}
or, 
\begin{eqnarray}
dU_{t} &=&\left\{ \left( S_{ab}-\delta _{ab}\right) d\Lambda _{ab}+\left(
S_{ab}-\delta _{ab}\right) \beta _{a}^{\ast } \, dB_{b}\right. \nonumber \\
&&\left. +\left( S_{ab}-\delta _{ab}\right) \beta _{b}\, dB_{a}^{\ast }+\left(
S_{ab}-\delta _{ab}\right) \beta _{a}^{\ast }\beta _{b} \, dt \right\} U_{t} \nonumber \\
&&-iHU_{t}dt.
\label{eq:UQSDE}
\end{eqnarray}
We now obtain the position and momentum QSDEs 
\begin{eqnarray*}
\dot{q}_{t} &=&\frac{1}{m}p_{t}, \\
\dot{p}_{t} &=&-V^{\prime }\left( q_{t}\right) \\
&& -i\hbar (b_{a}^{\ast }\left(
t\right) +\beta _{a}^{\ast }(t))\left( S_{ca}^{\ast }S_{cb}^{\prime }\right)
(b_{b}\left( t\right) +\beta _{b}\left( t\right) ).
\end{eqnarray*}
The averages now lead to 
\begin{eqnarray*}
\frac{d}{dt}\left\langle q_{t}\right\rangle &=&\frac{1}{m}\left\langle
p_{t}\right\rangle , \\
\frac{d}{dt}\left\langle p_{t}\right\rangle &=&-\left\langle V^{\prime
}\left( q_{t}\right) \right\rangle -i\hbar \beta _{a}^{\ast }(t)\left(
S_{ca}^{\ast }S_{cb}^{\prime }\right) \beta _{b}\left( t\right) .
\end{eqnarray*}

In the special case of a mirror, where we have only an input $b=b_{r}$ from
the right (say) then the equation (\ref{eq:QSDE_S}) above simplifies to 
\begin{eqnarray}
\dot{q}_{t} &=&\frac{1}{m}p_{t},  \notag \\
\dot{p}_{t} &=&-V^{\prime }\left( q_{t}\right) +\hbar b^{\ast }\left(
t\right) \theta ^{\prime }\left( q_{t}\right) b\left( t\right) .
\label{eq:QSDE_R}
\end{eqnarray}
where $\beta $ is the intensity of the coherent input field from the right,
and 
\begin{eqnarray*}
S=e^{i\theta \left( q\right) }
\end{eqnarray*}
is the reflection coefficient.

\begin{figure}[hbp]
	\centering
		\includegraphics[width=0.4500\textwidth]{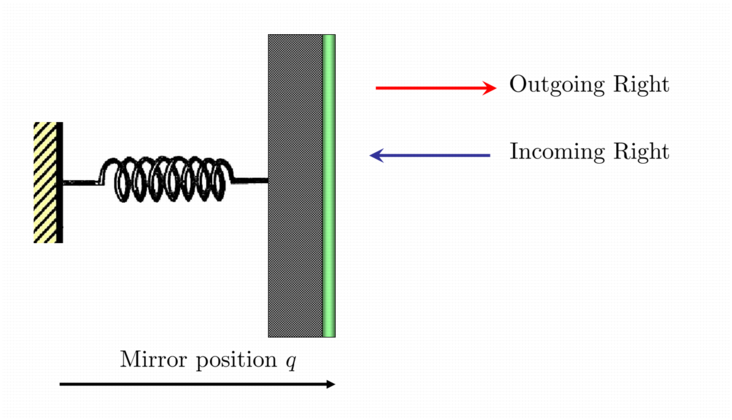}
	\caption{(Color online) Optomechanical model where an imperfect mirror with quantum mechanical position $q$ is in interaction with an input field.}
	\label{fig:Jump_opto}
\end{figure}

\section{Examples}

\subsection{Mirrors}

In the case of a perfect mirror at position $q$ we have form (\ref
{eq:perfect_mirror}) $R=e^{-i2kq}$ and so (\ref{eq:QSDE_R}) yields 
\begin{eqnarray*}
\dot{p}_{t}=-V^{\prime }\left( q_{t}\right) -2\hbar k\,b^{\ast }\left(
t\right) b\left( t\right) .
\end{eqnarray*}
This has the natural interpretation that the forcing term is the mechanical
force due to the potential $V$ and the radiation pressure which is the de
Broglie momentum $\hbar k$ of the photon (doubled due to the reflection)
times the number intensity $b^{\ast }\left( t\right) b\left( t\right) $. In
a coherent state of intensity $\beta $, this yields 
\begin{eqnarray*}
\frac{d}{dt}\left\langle p_{t}\right\rangle =-\left\langle V^{\prime }\left(
q_{t}\right) \right\rangle -2\hbar k\left| \beta (t)\right| ^{2}.
\end{eqnarray*}

In the case of a singular dielectric boundary at $q$ we obtain from (\ref
{eq:R_sdb}) 
\begin{multline*}
\dot{p}_{t}=-V^{\prime }\left( q_{t}\right) \\
-2\hbar k\,b^{\ast }\left( t\right) \frac{\mu ^{2}k^{2}\left( \cos
2kq_{t}-1\right) +2\mu k\sin \left( 2kq_{t}\right) }{\mu ^{2}k^{2}\left(
\cos 2kq_{t}-1\right) -2\mu k\left( \sin 2kq_{t}-1\right) }b\left( t\right) ,
\end{multline*}
which reduces to the perfect mirror expression in the limit $\mu \rightarrow
\infty $.

\subsection{A Dielectric Particle}

We may consider a point particle of mass $m$ with dielectric strength $\mu $%
. The scattering matrix for photons of wave vector $k$ is therefore from (%
\ref{eq:S_singular_point}) 
\begin{eqnarray*}
S=\frac{1}{1-\frac{i}{2}\mu k}\left[ 
\begin{array}{cc}
\frac{i}{2}\mu ke^{-2ikq} & 1 \\ 
1 & -\frac{i}{2}\mu ke^{2ikq}
\end{array}
\right]
\end{eqnarray*}
and we have 
\begin{eqnarray*}
S^{\ast }S^{\prime }=\frac{k}{1+\frac{1}{4}\mu ^{2}k^{2}}\left[ 
\begin{array}{cc}
-\frac{i}{2}\mu ^{2}k^{2} & \mu ke^{2ikq} \\ 
\mu ke^{-2ikq} & \frac{i}{2}\mu ^{2}k^{2}
\end{array}
\right] .
\end{eqnarray*}
The Langevin equation for $p_{t}$ is in this case 
\begin{eqnarray*}
\frac{d}{dt}p_{t} &=&-V^{\prime }\left( q_{t}\right) \\
&&-\frac{\hbar k}{1+\frac{1}{4}\mu ^{2}k^{2}}\left\{ \frac{1}{2}\mu ^{2}k^{2}%
\left[ b_{r}^{\ast }(t)b_{r}(t)-b_{l}^{\ast }(t)b_{l}(t)\right] \right. \\
&&\left. +i\mu kb_{r}^{\ast }(t)e^{2ikq_{t}}b_{l}(t)+i\mu kb_{l}^{\ast
}(t)e^{-2ikq_{t}}b_{r}(t)\right\} .
\end{eqnarray*}

In the limit $\mu \rightarrow \infty $ of infinite dielectric constant we obtain 
\begin{eqnarray*}
\frac{d}{dt}p_{t}=-V^{\prime }\left( q_{t}\right) -2\hbar k\left[
b_{r}^{\ast }(t)b_{r}(t)-b_{l}^{\ast }(t)b_{l}(t)\right]
\end{eqnarray*}
consistent with a perfect two-sided mirror at $q_{t}$ with left and right
field quanta reflected with momenta $\pm \hbar k$ respectively.

\subsection{The Adiabatic Elimination Example}

We have the reflection coefficient $R=S$ given by (\ref{eq:S_AE}) so that 
\begin{eqnarray*}
\frac{d}{dt}p_{t} &=&-V^{\prime }\left( q_{t}\right)  \\
&&-4\hbar k\,b^{\ast }\left( t\right) \frac{\gamma \Delta g_{0}^{2}\cos
\left( 2kq_{t}\right) }{\gamma ^{2}\Delta ^{2}+4g_{0}^{4}\cos ^{4}\left(
kq_{t}\right) }b\left( t\right) .
\end{eqnarray*}

\section{Quantum Measurement}

We now turn to the filtering problem, namely how do we best estimate the
state of the mirror from observations of the reflected output fields. We
consider a detector located in the far zone where we measure some observable 
$Y\left( t\right) $ of the field at time $t$. The time to go from the mirror
to the detector will be assumed negligible. The set of observables $\left\{
Y\left( s\right) :0\leq s\leq t\right\} $ is assumed to be commutative, and
our aim is to calculate the conditional density matrix $\varrho _{t}$ based
on these observations \cite{Belavkin,Carmichael}.

The most flexible approach is to use the use the theory of quantum
filtering. Let $\mathfrak{Y}_{t]}$ be the von Neumann algebra generated by $%
\left\{ Y\left( s\right) :0\leq s\leq t\right\} $. The we aim to compute,
for each system operator $X$ the conditional expectation 
\begin{equation*}
\widehat{X}_{t}=\mathbb{E}\left[ j_{t}\left( X\right) |\mathfrak{Y}_{t]}\right] 
\end{equation*}
of the Heisenberg picture value of the operator at time $t$ onto the algebra
generated by the measurements up to that time. We shall use established
results to derive explicit dynamical equations for $\pi _{t}\left( X\right) $
and therefore, through the identification 
\begin{equation*}
\widehat{X}_{t}=tr\left[ \varrho _{t}X\right] ,
\end{equation*}
for $\varrho $ itself. We shall use the filtering equations derived in 
\cite{GK} for coherent state inputs. We recall that
the SLH triple in this problem 
\begin{equation*}
S=\left[ 
\begin{array}{cc}
S_{rr} & S_{rl} \\ 
S_{lr} & S_{ll}
\end{array}
\right] ,
\end{equation*}
with the components dependent on the observable $q$, and $L$ vanishing. The filtering problem in the presence of fields in
coherent states with amplitudes $\beta _{r}(t)$ and $\beta _{l}\left(
t\right) $ respectively, is then equivalent to the vacuum filtering problem
with non-zero coupling 
\begin{equation*}
L^{\beta \left( t\right) }=S\left[ 
\begin{array}{c}
\beta _{r}\left( t\right)  \\ 
\beta _{l}\left( t\right) 
\end{array}
\right] .
\end{equation*}
This, of course, is explicitly contained in the unitary QSDE (\ref{eq:UQSDE}).
Note that $L^{\beta \left( t\right) \dag }L^{\beta \left( t\right) }=\left\|
\beta \left( t\right) \right\| ^{2}$ where we have the norm $\left\| \beta \left( t\right)
\right\| ^{2}=\left| \beta _{r}\left( t\right) \right| ^{2}+\left| \beta
_{l}\left( t\right) \right| ^{2}$.

\subsection{Homodyne Measurement}

Let $B_{\text{in},k}\left( t\right) =\int_{0}^{t}b_{k}\left( s\right) ds$ be
the input annihilation process, then we might aim to measure the fields 
\begin{equation*}
Y_{a}(t)=U\left( t\right) ^{\dag }\,\left[ 1\otimes \left( B_{\text{in}%
,a}\left( t\right) +B_{\text{in},a}\left( t\right) ^{\dag }\right) \right]
\,U\left( t\right) 
\end{equation*}
which gives the output quadrature for $a=l,r$. It follows from the quantum It%
\={o} calculus that 
\begin{equation*}
dY_{a}(t)=\sum_{b=l,r}j_{t}\left( S_{ab}\right) \left[ dB_{\text{in}%
,b}\left( t\right) +\beta _{b}\left( t\right) dt\right] +\text{H.c.}
\end{equation*}
The process is a diffusion with $\left( dY_{a}\right) ^{2}=dt$.

The filter equation for homodyne measurement is then, from equation (20) of 
\cite{GK}, 
\begin{multline*}
d\widehat{X}_{t} = \widehat{\mathcal{L}X}_{t}\,dt \\
+\sum_{a}\left\{ \sum_{b} \left( \widehat{XS_{ab}} \right)_t \beta _{b}\left( t\right)
+\sum_{b} \left( \widehat{S_{ab}^{\dag }X} \right)_t \beta _{b}^{\ast } \right. \\
\left.
 -\widehat{X}
_{t}\,\lambda _{a}(t)\right\} \,dI_{a}\left( t\right) 
\end{multline*}
where 
\begin{eqnarray*}
\mathcal{L}\left( X\right)  &=&\sum_{a,b}\beta _{a}^{\ast }\left( t\right) %
\left[ \sum_{c}S_{ac}^{\dag }XS_{cb}-\delta _{ab}X\right] \beta _{c}\left(
t\right)\\
&& -\frac{i}{\hbar }\left[ X,H\right] , \\
\lambda _{a}(t) &=&\sum_{b}\left( \widehat{S_{ab}} \right)_t \beta _{b}\left( t\right)
+\sum_{b}\left( \widehat{S_{ab}^{\dag }}\right)_t \beta _{b}^{\ast }
\end{eqnarray*}
and 
\begin{equation*}
dI_{a}\left( t\right) =dY_{a}\left( t\right) -\lambda _{a}\left( t\right)
\,dt.
\end{equation*}
The processes $I_{a}$ are independent Wiener processes.

The corresponding stochastic master equation is then
\begin{equation*}
d\varrho _{t}=\mathcal{D}\varrho _{t}\,dt+\sum_{a}\mathcal{H}_{a}\left[
\varrho _{t}\right] \,dI_{a}\left( t\right) ,
\end{equation*}
with
\begin{eqnarray*}
\mathcal{D}\varrho  &=&\sum_{a,b,c}S_{ab}\varrho S_{ac}^{\dag }\,\beta
_{b}\beta _{c}^{\ast }-\left\| \beta \left( t\right) \right\| ^{2}\varrho +%
\frac{i}{\hbar }\left[ \varrho ,H\right] , \\
\mathcal{H}_{a}\left[ \varrho \right]  &=&S_{ab}\varrho \beta _{b}+\varrho
S_{ab}^{\dag }\beta _{b}^{\ast }-\lambda _{a}\left( t\right) \varrho .
\end{eqnarray*}
We note that the mapping $\mathcal{H}_{a}$ is nonlinear in $\varrho$ since $\lambda _{a}\left( t\right)
=\sum_{b}tr\left\{ \varrho _{t}\left( S_{ab}\beta _{b}\left( t\right)
+S_{ab}^{\dag }\beta _{b}^{\ast }(t) \right) \right\} $.

In the case where there
is only one input (say the right side), we have $S=e^{i\theta \left(
q\right) }$ and the stochastic master equation simplifies to
\begin{eqnarray*}
d\varrho _{t} &=&\left\{ e^{i\theta }\varrho _{t}e^{-i\theta }-\varrho _{t}+%
\frac{i}{\hbar }\left[ \varrho _{t},H\right] \right\} dt \\
&&+\left\{ e^{i\theta }\beta \left( t\right) \varrho _{t}+\varrho
_{t}e^{-i\theta }\beta \left( t\right) ^{\ast }-\lambda \left( t\right)
\varrho _{t}\right\} dI\left( t\right) 
\end{eqnarray*}
with $\lambda \left( t\right) =tr\left\{ \varrho _{t}\left( e^{i\theta
}\beta \left( t\right) +e^{-i\theta }\beta ^{\ast } (t) \right) \right\} $%
.

\subsection{Photon Counting}

Now set $\Lambda _{\text{in},a}\left( t\right) =\int_{0}^{t}b_{a}^{\ast
}\left( s\right) b_{a}\left( s\right) ds$ be the input annihilation process,
then we might aim to measure the fields 
\begin{equation*}
Y_{a}(t)=U\left( t\right) ^{\dag }\,\left[ 1\otimes \Lambda _{\text{in}%
,a}\left( t\right) \right] \,U\left( t\right) 
\end{equation*}
which gives the output quadrature for $a=l,r$. The $Y_{a}$ are
(time-inhomogeneous) Poisson processes.

The filter equation for photon counting measurement is then, from equation (21) of 
\cite{GK}, 
\begin{eqnarray*}
&&d\widehat{X}_{t} =\widehat{\mathcal{L}X}_{t}\,dt \\
&+&\sum_{a}\left\{ \frac{1}{\nu _{a} (t)}
 \sum_{b,c} ( \widehat{S_{ab}^{\dag }XS_{ac}  } )_{t}
\,
\beta _{b}^{\ast }\left( t\right) \beta _{c}\left(
t\right) -\widehat{X}_{t} \right\}   dJ_{a}\left( t\right) 
\end{eqnarray*}
where $\mathcal{L}\left( X\right) $ is as before, and
\begin{equation*}
\nu _{a}(t)=\sum_{b,c} \left( \widehat{S_{ab}^{\dag }S_{ac}}\right)_t \beta _{b}^{\ast
}\left( t\right) \beta _{c}\left( t\right) 
\end{equation*}
and 
\begin{equation*}
dJ_{a}\left( t\right) =dY_{a}\left( t\right) -\nu _{a}\left( t\right) \,dt.
\end{equation*}
From the unitarity of $S$, we have the identity
\begin{equation*}
\sum_a \nu _{a}(t)= \| \beta (t) \|^2 , 
\end{equation*}
and this allows us to write

\begin{eqnarray*}
&& d\widehat{X}_{t} = \dfrac{1}{i\hbar} \widehat{ [X,H]}_t \, dt\\
&& +\sum_{a}\left\{ \frac{1}{\nu _{a} (t)}
 \sum_{b,c} ( \widehat{S_{ab}^{\dag }XS_{ac}  }  )_{t}
\,
\beta _{b}^{\ast }\left( t\right) \beta _{c}\left(
t\right) -\widehat{X}_{t} \right\}   dY_{a}\left( t\right) .
\end{eqnarray*}

The corresponding stochastic master equation is now
\begin{equation*}
d\varrho _{t}= \dfrac{1}{i\hbar} [H, \varrho _{t}] \,dt+\sum_{a}\mathcal{H}_{a}\left[
\varrho _{t}\right] \,dY_{a}\left( t\right) 
\end{equation*}
with 
\begin{equation*}
\mathcal{H}_{a}\left[ \varrho \right] =\frac{1}{\nu _{a}\left( t\right) }%
\sum_{b,c}S_{ac}\varrho S_{ab}^{\dag }\,\beta _{b}^{\ast }\left( t\right)
\beta _{c}\left( t\right) -\varrho ,
\end{equation*}
and $\nu _{a}\left( t\right) =tr\left\{ \varrho _{t}\sum_{b,c}S_{ab}^{\dag
}S_{ac}\right\} \beta _{b}^{\ast }\left( t\right) \beta _{c}\left( t\right) $%
. 

This, of course, corresponds to a continuous Hamiltonian evolution under $H$, with
jumps 
\begin{equation*}
\varrho \to \dfrac{1}{\nu _{a}\left( t\right) }%
\sum_{b,c} \beta _{c}\left( t\right) S_{ac}\varrho S_{ab}^{\dag }\,\beta _{b}^{\ast }\left( t\right)
\end{equation*}
occurring at random times when we detect a photon at the right ($a=r$) or left ($a=l$)
detector. 

Again, this simplifies if we have only one input, and we find the
stochastic master equation simplifies to
\begin{eqnarray*}
d\varrho _{t} &=&\left\{ e^{i\theta }\varrho _{t}e^{-i\theta }-\varrho _{t}+%
\frac{i}{\hbar }\left[ \varrho _{t},H\right] \right\} dt \\
&&+\left\{ e^{i\theta }\varrho _{t}e^{-i\theta }-\varrho _{t}\right\}
dJ\left( t\right)  \\
&=&\frac{i}{\hbar }\left[ \varrho _{t},H\right] \,dt+\left\{ e^{i\theta
}\varrho _{t}e^{-i\theta }-\varrho _{t}\right\} \,dY\left( t\right) ,
\end{eqnarray*}
and $\lambda \left( t\right) =\left| \beta \left( t\right) \right| ^{2}$.

\begin{acknowledgments}
The author wishes to thank the Isaac Newton Institute for Mathematical Sciences, 
Cambridge, for support and hospitality during the programme \textit{Quantum Control Engineering} 
where work on this paper was completed. He
acknowledges several fruitful discussions with Howard Wiseman, Andrew
Doherty, Ramon van Handel, Luc Bouten, Hendra Nurdin, Jake Taylor and
Matthew James.
\end{acknowledgments}

\appendix

\section{The Holevo Time-Ordering and}

We now justify the limit of $\tilde{U}( t) $ as a Holevo time-ordered
exponential 
\begin{widetext}

\begin{eqnarray}
\tilde{U}^{( n) }( t+\tau ,t) =\tilde{U}^{(
n) }( t+\tau ) \tilde{U}^{( n) }( t)
^{\dag }=\sum_{r=0}^{n}( -i) ^{r}\int_{t+\tau \geq t_{r}>\cdots
t_{1}\geq 0}\tilde{H}^{( n) }( t_{r}) \cdots \tilde{H}
^{( n) }( t_{1}) dt_{r}\cdots dt_{1}.
\label{eq:app1}
\end{eqnarray}
For a fixed $n$, we partition the interval $[ t,t+\tau ] $ with
grid points $\sigma _{j}=t+\frac{j\tau }{N}$ for $j=1,\cdots ,N$ where 
$N=\frac{n\tau }{2c}$. Each of the first $N$ terms in the expansion of (\ref{eq:app1}) may be
approximated as the discrete sums
\begin{eqnarray}
( -i) ^{r}\sum_{N\geq j_{r}\geq \cdots \geq j_{1}\geq 1}\tilde{H}
^{( n) }( \sigma _{j_{r}}) \cdots \tilde{H}^{(
n) }( \sigma _{j_{1}}) .
\label{eq:app2}
\end{eqnarray}
Now we compare this to the exponential $\exp \{ -i\sum_{j=1}^{N}\tilde{H
}^{( n) }( \sigma _{j}) \} $ which we may
likewise expand leading to the $r$th term
\begin{eqnarray}
\frac{ ( -i )^{r}}{r!}
\sum_{j_{r}, \cdots , j_1=1}^{N} 
\tilde{H}^{( n) } ( \sigma _{j_{r}}) 
\cdots \tilde{H}^{( n) }( \sigma _{j_{1}}) =
\frac{ ( -i ) ^{r}}{r!}
\sum_{j_r, \cdots , j_1=1}^{N} E_{\alpha_{ j_{r}} \beta_{ j_r} }\cdots E_{\alpha_{ j_1} \beta_{j_1 }}
\otimes \tilde{\lambda}_{\alpha_{j_{r}} \beta_{ j_{r}} }^{( n) }( \sigma
_{j_{r}}) \cdots \tilde{\lambda}_{\alpha_{ j_{1}} \beta_{ j_{1} }}^{( n) }( \sigma _{j_{1}}) .
\label{eq:app3}
\end{eqnarray}
\end{widetext}
The crucial observation is that $\tilde{\lambda}_{\alpha \beta }^{( n) }( t) 
$ and $\tilde{\lambda}_{\mu \nu }^{( n) }( s) $ will commute whenever $|
t-s| \geq \frac{c}{n}$, and so for $r\leq N$ the various $\tilde{\lambda}^{(
n) }$ terms commute in the expression above. Therefore for a fixed set of
indices $\alpha_{ j_{r}} ,\beta_{ j_{r} } ,\cdots \alpha_{j_{1} }
,\beta_{j_{1}} $ we may reorder the $\tilde{\lambda}^{( n) }$'s in (\ref{eq:app3}%
) in $r!$ equivalent ways, thereby recovering (\ref{eq:app2}). Therefore the
first $N$ terms of series expansion (\ref{eq:app1}) agree with the first $N$
terms of $\exp \{ -i\sum_{j=1}^{N}\tilde{H}^{( n) }(\sigma _{j}) \} $.

We then take the limit $n\to \infty $ and $\tau \to 0$ to obtain in
principle the Holevo time-ordered exponential (\ref{eq:Holevo}).

\end{document}